# Einstein's Unpublished Opening Lecture for His Course on Relativity Theory in Argentina, 1925

Alejandro Gangui

*IAFE/Conicet and CEFIEC, FCEyN-UBA, Buenos Aires*

Eduardo L. Ortiz

*Imperial College, London*

## Introduction

In 1922 the University of Buenos Aires (UBA) Council approved a motion to send an invitation to Albert Einstein to visit Argentina and give a course of lectures on his theory of relativity. The motion was proposed by Jorge Duclout (1856-1927), who had been educated at the Eidgenössische Technische Hochschule, Zürich (ETH). This proposal was the culmination of a series of initiatives of various Argentine intellectuals interested in the theory of relativity. In a very short time Dr. Mauricio Nirenstein (1877-1935), then the university's administrative secretary, fulfilled all the requirements for the university's invitation to be endorsed and delivered to the sage in Berlin. The visit took place three years later, in March-April 1925.

The Argentine press received Einstein with great interest and respect; his early exchanges covered a wide range of topics, including international politics and Jewish matters. Naturally, the journalists were more eager to hear from the eminent pacifist than from the incomprehensible physicist. However, after his initial openness with the press, the situation changed and Einstein restricted his public discourse to topics on theoretical physics, avoiding some controversial political, religious, or philosophical matters that he had freely touched upon in earlier interviews.

Immediately after Einstein's visit, Mauricio Nirenstein published a short paper in which he made some interesting remarks on the visit and the visitor. His paper was presented as a dialogue, a personal conversation in which Einstein outlined his views on the epistemology of the physical sciences. In a footnote, Nirenstein explained that Einstein's thoughts on epistemology were based on an unpublished text. However, in his reference to Einstein's text on epistemology, he indicated that Einstein's remarks were given in response to a recent newspaper article. He did not mention that Einstein, in fact, had written this text for his opening lecture at the UBA. Consequently, he did not explain why he did not publish it verbatim or in translation.

Einstein's unpublished piece on the epistemology of physics became known in Argentina as the *discurso inédito* and it was rumored that Einstein had left it in the hands of Nirenstein. Six years after the visit, in 1931, an avant-gard literary journal, *La Vida Literaria*, published a Spanish translation of the *inédito* as well as photographs of two fragments of the manuscript, which serves to test the quality of the Spanish translation and shows that it was addressed to the UBA audience. The note in *La Vida Literaria* also referred briefly to Einstein's encounters with the local press and to the impact these meetings may have had on his decision not to use the *inédito* as an introduction to his set of lectures.

The English translation of the *inédito* is based on three sources: the Spanish translation of the German text, written by Baldomero Sanín Cano and published in *La Vida Literaria*, Buenos Aires, 1931, which covers the whole text; photographs of pages 1 and 5 of the manuscript in German in Einstein's hand, reproduced in the same journal; and transcribed excerpts from Einstein's manuscript printed in J. A. Stargardt's catalogs number 615 (1978), 117, and number 683 (2006), 188.

This introduction gives a brief account of the individuals, institutions, and events related to the *inédito* and compares it with the text of several lectures given by Einstein in Argentina where he briefly considered matters related to the philosophy of science. Einstein's lecture may not rank among the best of Einstein's pieces on epistemology, but it is, nevertheless, a succinct and candid account of his views on the subject in the year 1925. For the background and details on Einstein's visit to Argentina and its impact there and in Uruguay, the reader is referred to the monograph Ortiz 1995.

**Holmberg's Circle and Mauricio Nirenstein**

Dr. Eduardo L. Holmberg (1852-1937) came from a patrician Argentine family with deep roots in the country; in the early twentieth century he was regarded as one of Argentina's finest writers of his generation and credited with the introduction of fantastic literature in his country.[1] In his novels, Holmberg employed modern scientific ideas, in particular, Darwinism, which he as a scientist had helped to introduce in Argentina as early as the 1870s. A medical doctor by training and a naturalist by vocation, he was a professor of Natural Sciences at the UBA. In 1925, as President of the National Academy of Exact, Physical, and Natural Sciences, Buenos Aires, it was Holmberg who suggested that Einstein should be honored by the Academy and invited to participate in a special session on relativity theory.

Toward the end of the nineteenth century young writers and poets became known through readings of their works at Holmberg's literary salon where, in the spirit of "modernism," they freely mixed with scientists. Mauricio Nirenstein was one of the young members of that circle. Nirenstein was born in Egypt of Jewish parents in 1877 and was taken to Argentina in 1889, when foreign immigration was strongly encouraged there. He studied law at the UBA but also showed an interest in literature, contributing to local magazines with short essays, stories, and poetry. His literary work had limited originality; however, his contact with Holmberg, who was interested in Germanic literature and had written on Goethe, launched Nirenstein into a study of German literature.

---

[1] This topic is considered in some detail in Ortiz 2005.

2
07/03/2009

Later, in parallel with school and university teaching, Nirenstein had a long and successful career at the administrative offices of the UBA, which he joined as a clerk in 1897, became pro-Secretary in 1906, and was finally designated University Secretary in 1922; he kept this position until his retirement in 1930. At the UBA he lectured on northern European literature (at the Faculty of Humanities) and also on political economics (at the Faculty of Economics).

Nirenstein belongs to the first generation of Argentine-Jewish young professionals, artists, and intellectuals who began to integrate more fully into Argentina's cultural movements. The foundation in the early 1920s of a cultural institution called Asociación Hebraica Argentina (later called Sociedad Hebraica Argentina, SHA) testifies to their wish to project an image of their Jewish identity. Leaders of the SHA wished to inaugurate their association's activities by bringing the most eminent representative of a new generation of European-Jewish intellectuals who, besides being outstanding in science, shared their pacifist and advanced social and political ideals; they thought of Einstein.

**University Reform and the University of Buenos Aires' Invitation**
In 1918 an influential movement of university reform took Argentine universities by storm (see Del Mazo 1955; Ciria and Sanguinetti 1983). This movement was a direct response of Argentine students to the impact of the Mexican revolution of 1910, the Russian revolution of 1917, and the political convulsions that followed the end of World War I in Europe. The Reform movement helped to modernize the university structure, gave students some voice in academic affairs, and suggested the development of disciplines with a more direct impact on Argentina's contemporaneous technical, economical, and social problems. With the added pressure of a first generation of young sons of immigrants, who had benefited from a free, non-clerical, and well developed education system reaching up to university level, the Reform movement opened the ranks of university teachers to a wide social spectrum.

Nevertheless, while influential at home, and also in Latin America in general, the Reform movement was short lived. In the early 1920s, prosperity began to shine again in food producing Argentina and its economy began to return to its thriving state of the prewar years. Gradually, though with some concessions, less idealistic authorities tried to return Argentine political and university life to its traditional path. To achieve that objective they even resorted occasionally to the use of violence. In 1930 a military uprising toppled the elected government and the more conservative university authorities began to adapt to the realities of world depression (Halperin Donghi 2003).

In 1922, the same year Nirenstein was designated the UBA Secretary, the university began considering proposals to invite Einstein to Argentina. Einstein's invitation was very much in tune with the pacifist ideals of the Reform movement, which Duclout, one of the earliest advocates of relativity theory in Argentina, fully shared. Due to that convergence of objectives, members of the SHA also played a role in the gestation of the invitation, offering to help draft some wealthy German-Jewish businessmen established in Argentina to contribute financially and make the invitation more attractive. It was finally arranged that the visit would take place in March-April 1925.

Einstein's invitation to visit Argentina was part of a much wider and persistent effort to attract foreign university teachers to help modernize Argentina's scientific and technical base, a process the country had started in the 1870s and intensified further after the end of the First World War. In the same year Einstein was invited to visit Argentina, the eminent Berlin professor Georg-Friedrich Nicolai (1874-1964), a well known scientist and pacifist (Zuelzer 1981), as well as one of Einstein's personal friends and co-signatory of the counter-manifest of 1914, accepted the offer of the chair of physiology at the old University of Córdoba, in Central Argentina (Ortiz 1995). The latter, founded in the seventeenth century as a Jesuit school, was also the place where the Reform movement had started precisely four years before Nicolai's arrival.

**The Administration of the Visit**
In March 1925, as planned, the ship on which Einstein traveled to Argentina approached the coast of South America. The UBA's Council sent a delegation of Argentine academics, which included its administrative secretary Nirenstein, to Montevideo, its last stop before reaching Buenos Aires. They welcomed Einstein there and accompanied him on the last leg of his journey across the River Plate to Buenos Aires. As a German speaker and a man intimately linked to the UBA and the SHA, and also in touch with the main financial contributors to the visit and with the German-Argentine Jewish community, Nirenstein was a natural choice as Einstein's personal contact while in Argentina. In addition, he was a tactful, prudent, and well connected man who sensed the mood of national politics as well as that of the different groups sponsoring the visit.

From that journey on he established friendly contact with Einstein and later played a role of some influence in managing his visit. Sometimes with his wife, one of Holmberg's daughters, Nirenstein accompanied Einstein on his academic trips inside Argentina. He would not, however, qualify to accompany Einstein to the *estancias* or to the exclusive parties offered by the wealthy donors, such as Mr. B. Wassermann or Mr. A. Hirsch.

Einstein opened his contact with Argentina with a rich agenda in which he displayed the remarkable free tone he then used in Berlin (Scheideler and Goenner 1997; Renn and Sauer 1997; Levenson 2003; Rowe and Schulmann 2007). From his newspaper interviews on arrival in Montevideo, and immediately after he landed in Buenos Aires, it became clear that he was prepared to enter into discussion of serious topics of science, but also of matters well outside science. He offered his views on relativity, on physics, and on science in general, but was also willing to enter into discussion on the state of science and culture in Europe after the War, on pacifism and militarism, on the end of pogroms and the changing perspectives for Jews in Russia after the Revolution, on Zionism and assimilation; on the question of choosing between Hebrew and Yiddish as the Jewish national language, and on several other equally interesting topics.

Besides, when the leading local newspaper *La Prensa* asked him for an article to be published immediately before his arrival in Buenos Aires, Einstein sent one whose subject was neither relativity, nor physics or science, but *Pan-Europeanism* (Einstein 1925a). In his article Einstein developed his ideas on the creation of a truly European movement of economic, political, and cultural integration that might avoid repeating the mistakes that led to the catastrophe of the First World War.

This agenda, as well as his contacts with the press, changed a few days after his arrival in Buenos Aires. It is difficult to know precisely why, but there are several indicators which suggest he may have been advised to refrain from eloquence on political, religious, philosophical, or sensitive matters and confine his exchanges more specifically to the field of relativity theory.

The intense political antagonism in the Germany of these years was fully reflected in the still powerful German community in Argentina (Newton 1977). The cool reception the German embassy organized for him in Buenos Aires is a testimony of such antagonism. Einstein perceived this as offensive and reacted with rage.[2] Einstein's pacifist past had even made the German physicist Richard Gans, then director of the main physics institute in Argentina, hesitate when Einstein's invitation to visit Argentina was discussed at the enlightened German-Argentine cultural association; he supported an alternative name. However, the association finally joined others in granting financial and intellectual sponsorship to the visit.

It is also a fact that in Argentina, as indeed in the world, the atmosphere of the time when the invitation was issued had changed substantially by 1925. Idealistic and egalitarian movements, such as university Reform, were in clear retreat in Argentina by 1925. Even Nicolai was already facing difficulties at the University of Córdoba by the time Einstein arrived in Argentina, and was pressed to leave soon after.

---

[2] Einstein wrote in his *Diary*: "Droll people these Germans. I am a stinking flower to them and nevertheless again and again they use me in their button-hole" (quoted in Ortiz 1995, 114).

In light of the revolutionary changes of his relativity theory and his standing as a pacifist, Einstein's contacts with groups of progressive intellectuals who passionately supported his visit to Argentina were quite limited. In his influential *Revista de Filosofía*, José Ingenieros (1877-1925)[3] saluted the arrival of Einstein to Argentina as a landmark in the history of the country's culture; the journal also started serializing his university lectures. In the past, *Revista de Filosofía* had published interesting notes on relativity theory which may have helped to stimulate the interest of UBA's council members on the subject. Nevertheless, during his visit, public contacts between Einstein and members of this intellectually vibrant group were rare.

His visit had great public and academic impact, with warm response from university students and enthusiastic support from popular sections of the Argentine-Jewish community in various Argentine cities. However, there was only limited, polite acknowledgment of his presence in more traditional sections of Argentine society, whose hegemony was again in ascent after the impact of the World War I began to fade. Their response was much less receptive to him than it had been to the frequent visits of foreign writers, artists, or savants, attracted by the country's economic prosperity since the early 1870s.

One might argue that Nirenstein, as a close confidant, may have been aware of the effort to steer the visit in a direction which he, and possibly an important group of those who contributed to sponsoring and financing the visit, interpreted as more in tune with the results they desired. A dedication of a photograph given to Nirenstein by Einstein at the end of his journey, whose text is quoted in his *Diary*, suggests that Nirenstein had in fact helped to steer Einstein's visit "so that nobody has a chance to get angry."[4] Finally, as Einstein crossed to Uruguay, his *Diary* displays a sense of relief from the pressures and strain he experienced in Argentina.

**Einstein's Introduction to the Series of Lectures on Relativity Theory**

---

[3] Ingenieros, a fine writer, was an international figure in the field of psychology and one of Argentina's most respected, and also controversial, intellectuals of his time. In 1922, the year Einstein was invited to lecture in Argentina, his textbook on psychology was translated into German (*Prinzipien der Biologischen Psychologie*) on the advice of and with an introduction by Wilhelm Ostwald (see Ingenieros 1922).

[4] Einstein wrote in his *Diary*: "Well guided by your hand/ I groped bravely through this land/ he looks into everybody's heart/ so that nobody has a chance to get angry [Einstein used the Yiddish word *broges*]/ that is Professor Nierenstein" (quoted in Ortiz 1995, 115). Einstein mistakenly wrote Nierenstein for Nirenstein.

Let us briefly consider now the introductory lectures delivered by Einstein in Argentina. As we have indicated above, Einstein had written a more epistemologically oriented opening lecture which was never delivered; later in this paper we will contrast his introductory lectures at the UBA and at other places with the text he did not use, the *discurso inédito*.

In presenting his course on relativity theory at the UBA, Einstein gave two introductory lectures: a formal one in front of a wide and heterogeneous university audience, and a more technical one at the start of his specific course at the UBA's Faculty of Science. In both opening lectures he emphasized physics and physical experiments over epistemology.

The first lecture was given on Friday March 27 to a general public of teachers and students, in the presence of ministers of state and foreign ambassadors; it was widely reported in Argentine leading newspapers (Einstein 1925b). This lecture was brief, lasting for only about half an hour; in it Einstein sketched the fundamental principles of Galileo's mechanics, highlighting the role played by inertial reference systems and by the laws of inertia in that formulation. Immediately after he discussed briefly the physical experiments performed to verify the validity of the so-called principle of relativity of classical mechanics, and the need to rely on extra-mechanical considerations, he went on to consider Fizeau's experiments and finished by pointing at the difficulties created by notions of absolute space or æther.

The second inaugural lecture, given on the following day at the Faculty of Science, marked the beginning of his course in earnest; again, the leading national newspapers reported it in detail (Einstein 1925c). In this second lecture Einstein discussed in greater detail the delicate physical experiments carried out to verify the hypothesis of the existence of the æther and the nature of the difficulties encountered. He then considered the ingenious suggestions proposed by Lorentz and FitzGerald to overcome those difficulties. The lecture ended with a reference to the work of Michelson, indicating that by using methods taken from optics it was possible to design experimental techniques that allowed for a higher precision than was possible through mechanical devices. Einstein indicated that these results reinforced acceptance of the fact that the laws of optics must be independent from uniform translation movements of reference systems. He was then able to generalize the relativity principle of mechanics and to formulate the hypothesis of the constancy of the speed of light. The course of lectures continued with a mathematical exposition of relativity theory.

Coriolano Alberini (1886-1960) was a serious professional philosopher, deeply interested in bringing to Argentina the ideas of German contemporaneous philosophy, and a committed anti-positivist. At the time of Einstein's visit to Argentina he was Dean of the UBA's Faculty of Humanities and a powerful figure in the new more conservative world of post-Reform academia. He invited Einstein to open the academic courses for 1925 at his Faculty. There, Einstein discussed the impact of relativity theory on the concepts of space and time. A few days later he visited the University of Córdoba offering a shorter version of the same lecture.

If Alberini expected a distinctly "anti-positivist" philosophical lecture, he may have been slightly disappointed. Einstein's lecture at his Faculty started with a reference, in very general terms, to the views held by positivists and idealists on the origin of our conceptions and the role that experiment played in that process. He indicated that the positivist view was then prevalent among physicists and, instead of taking sides, he offered a humorous comparison of these two views. He moved immediately into a discussion on Euclidean geometry and physics, commenting on the impact that the theory of relativity had on them; in this last part he essentially followed the train of thought he had published in Einstein 1921.

Before Einstein's lecture, Alberini gave a presentation which he published in the Literary Supplement of *La Nación* on April 12 (Alberini 1925). Although he politely did not go over Einstein's lecture, Alberini insisted, nevertheless, on the anti-positivist character of Einstein's carefully constructed lecture. For a contemporaneous reader, Alberini's references to anti-positivism were a rebuff to the ideas of his colleague Ingenieros, a philosophical and political rival who was also a leading professor at the Faculty of Humanities.

**Nirenstein's Account of Einstein's Visit and the Elusive** *inédito*

Soon after Einstein left Argentina, in September 1925, a brief paper by Nirenstein entitled *Einstein en Buenos Aires* appeared in *Verbvm*, a journal edited by the Student Union (*Centro de Estudiantes de Filosofía y Letras*) of the UBA's Faculty of Humanities (Nirenstein 1925).

Nirenstein's paper contains some interesting insights into matters outside the realm of science. The paper begins with a section called "The Man"; the following section, "Zionist propaganda," is a humorous evaluation of how different groups in the Argentine-Jewish community perceived Einstein's visit (topics discussed were assimilation, which Nirenstein and the SHA organizers mainly supported, and Zionists). Nirenstein's impressions on the reaction of the media, particularly newspapers and newsreels, is discussed in the next two sections.

In the fifth section, Nirenstein compelled Einstein, who in any case had a good personal opinion of Alberini, to admit that had he been in Berlin the latter would have been made a member of the Prussian Academy of Sciences; needless to say Dean Alberini was Nirenstein's chief at the Faculty of Humanities. After a brief section on Goethe, the paper ends with an affectionate and interesting vignette on their visit to the home of Jorge Duclout, who studied at the same institution as Einstein in Zürich. At the time of the visit Duclout was

8
07/03/2009

ill and could not participate as much as he may have wished; Einstein, who respected him highly, made a point to visit him at his home.

In a footnote, Nirenstein indicated that the paper would include a section called "an epistemological dissertation" based on "an unpublished text by Einstein." Probably due to a printing error, such a section-title was not included. The epistemological discussion is, precisely, a selective review of the *inédito*. Instead of reproducing the notes faithfully, Nirenstein presented Einstein's text metaphorically, as a dialogue that took place during the return journey from Duclout's home, outside Buenos Aires, while Nirenstein was driving his car back to central Buenos Aires in the company of Einstein. That is, at an instance in the narrative when Einstein, visibly touched by the visit to Duclout, seemed prepared to enter into more intimate subjects.

Nirenstein remarked in his paper that "Now, as [Einstein] speaks in German, his expression becomes clear and neat and his precision, admirable," highlighting the communication problems Einstein faced while in Argentina, where he lectured in French which, he admitted, was far from perfect. Later, we will return to the epistemological notes in Nirenstein's dialogue.

**Samuel Glusberg and the Literary Magazine *La Vida Literaria***

The year 1931 marks a new stage in the history of Einstein's unpublished paper: a Spanish translation of the *inédito* appeared in the April issue of a limited circulation *avant-gard* literary magazine called *La Vida Literaria* (Einstein 1931).

*La Vida Literaria* is a rare journal; Harvard University library has microfilm copies of some of its issues. We located a set of originals in Buenos Aires, in a private collection to which we were given access in 2005. As we shall show, the text published in 1931 has many points in common with the "epistemological conversation" Nirenstein reported. Besides, some remarks added to the published text support further the notion that Einstein's free exchanges with the press had caused some concern.

*La Vida Literaria* was published in Buenos Aires between 1928 and 1931; it ceased publication soon after Einstein's *inédito* was printed. The journal was edited by the journalist and publisher Samuel Glusberg (1898-1987), born in Kishinev, Russia (Chişinău, Moldova), of Jewish parents and brought to Argentina in his early infancy, after the pogroms of 1903. Contributors to *La Vida Literaria* include some of the leading South American intellectuals of the time.[5]

---

[5] Such as writers Leónidas Barletta, Leopoldo Lugones, Ezequiel Martínez Estrada, Baldomero Sanín Cano, and Horacio Quiroga; philosopher Francisco Romero; art critic Leopoldo Hurtado, and illustrator and poet José Sebastián S. Tallon.

In the 1920s there were two main literary groups in literary Buenos Aires: the *realist*, called the *Boedo* group, and a vanguard one, called *Florida*.[6] The latter published the important literary journal *Martín Fierro*[7] to which young writers of the calibre of Jorge Luis Borges regularly contributed. These groups were united by a common interest in literary renewal, and although separated by different aesthetics, there were many bridges between them; Borges himself was one of them.

Although Glusberg did not hide his extreme left-wing views (he sponsored the literary and political journal *Extrema izquierda* and later embraced Leon Trotsky's view of socialism) he moved with ease among the different literary groups; no doubt, the fact that he was a publisher helped. Not only was he among the founders of *Martín Fierro*, but was also the editor of another literary journal, *Babel*, which took its name from Glusberg's own publishing house;[8] this last journal had a more radical edge than *Martín Fierro.* The list of authors in the catalog of Glusberg's publishing house, as well as the list of contributors to *Babel*, was predominantly left-wing, or at least unconventional, but both remained open to good writers, irrespective of their aesthetic views.

*Babel* stopped publication in 1928, only to reappear in the same year under the new name of *La Vida Literaria*. This metamorphosis may have been the consequence of a change in editorial policy.[9] In turn, the disappearance of *La Vida Literaria* in 1931 should be seen as a consequence of the climate created in Argentina by the military coup of 1930 we alluded to earlier; Glusberg had to go into exile in the mid-1930s.

A new reference to the unpublished lecture appeared in the Literary Supplement of the Argentine newspaper *La Nación*, in 1934, in a review of the French translation of Einstein's book *Mein Weltbild* written by the Argentine journalist Enrique Espinosa (Espinosa 1934). This reference brought news of the fact that Einstein's *inédito* had been published three years earlier in *La Vida Literaria* and reached a much larger circle of readers. Espinosa rightly stated that it had probably saved the text from being permanently lost. His reference informs us that the translation was the work of the Colombian writer and diplomat Baldomero Sanín Cano (1861-1957) who, by the time of Einstein's visit, resided in Buenos Aires as his country's Ambassador to Argentina and often contributed to the magazine with his own valuable literary work. By 1931, when the translation appeared in *La Vida Literaria*, Sanín Cano was still living in Argentina.

---

[6] The names are related to the street location of the *cafés* they frequented in the city of Buenos Aires.
[7] It takes its name from the *gaucho* who was hero of the celebrated Argentine epic poem by José Hernández (1834-1886).
[8] *Editorial Babel*, Buenos Aires.
[9] As Tarcus has suggested, the launching of a journal with a new name may have followed the dominance of a new trend of "Americanismo" (a tendency to return to the native civilizations of America) of which the Peruvian poet José Carlos Mariátegui (1894-1930) and the U.S. writer Waldo Frank (1899-1967) were then the leading representatives (Tarcus 2001).

The *inédito* continued to be a topic of interest in Argentina. A further reference appeared in 1980 in a note written by philosopher Diego F. Pró in the first volume of his edition of Alberini's correspondence (Alberini 1980, 99) where he essentially repeated what Espinosa had said in 1934, adding that the manuscript had been left in the hands of Nirenstein. Pró may not have had access to *La Vida Literaria*, since his reference to it is incorrect.

In 1955, in commemoration of Einstein's death, and with his name now slightly changed from Espinosa to Espinoza, the author published a note in *Davar*, the SHA's journal (Espinoza 1955). Its text is quite similar to that of his 1934 note. However, he confuses matters slightly stating now that the *inédito* had been published in *Babel*, not in its immediate successor, *La Vida Literaria*. He also adds that it was at his request that Sanín Cano translated the text from German, which is quite plausible and, without any further explanation, he added that "later [Einstein] omitted reading it" (ibid., 77). Espinoza's note is followed by a transcription of the *inédito* version published by *La Vida Literaria* in 1931 (Einstein 1955).

The 1931 issue of *La Vida Literaria* gives a far more explicit recounting of the events surrounding *inédito*. The front-page article states that Einstein's lecture was written "to serve as an introduction to his scientific lectures" in Argentina (see fig. 1 in the Appendix). More interestingly, it explains why Einstein did not use the written text in his university lectures:

> But, after the numerous statements local journalists managed to extract from him, the sage preferred to give his lectures directly [that is, without the epistemological introduction] at the Faculty of Engineering.[10]

This indicates the great impact that Einstein's public statements had on Argentina, as indeed elsewhere in the world including the United States.[11] Einstein's openness was well received in America, after having less success in Argentina. "The international impact of some of the papers he published there" finally broke Nirenstein's restraint and allowed the editors of *La Vida Literaria* to make the *discurso inédito* available to the very eager, but at the same time very exclusive, audience of that journal.

**Einstein's *inédito* in *La Vida Literaria***

From entries in Einstein's *Diary*, we know that while on board the *Cap Polonio*, Einstein mixed social life with reading philosophical and scientific literature. Émile Meyerson (1859-1933), the anti-positivist author of *La déduction relativiste* (Meyerson 1924) was one philosopher who moderately influenced him.[12] Meyerson's work was well known in

---

[10] The collective name used then for the engineering and science departments in the UBA was *Facultad de Ingeniería*; much later a *Facultad de Ciencias* became an independent entity.
[11] On Einstein's writings on politics, see Rowe and Schulmann 2007.
[12] He published a review of Meyerson's book (Einstein 1928). On the influence of Meyerson on Einstein's thought, see Zahar 1980.

Argentina at the time and he was invited to lecture in Buenos Aires the year before Einstein arrived in Argentina; however, Meyerson was unable to accept the invitation because of illness.

The fact that Einstein wrote his piece on the epistemology of the physical sciences with the idea of reading it in front of the university authorities is clear from its heading: "Honourable Rector, Honourable Professors, and Students of this University," which is reproduced in photograph in *La Vida Literaria* (see figs. 1, and 2 in the Appendix) and in Stargardt's catalog (see fig. 3 in the Appendix). In his written text Einstein expressed satisfaction on being invited to lecture in a "blessed land" where he could engage in a discussion with colleagues interested in clarifying matters of science at a time when "political and economic struggle and nationalistic fragmentation" were dividing the world. These interesting thoughts were neither included in Nirenstein's dialogue nor in any of Einstein's university public lectures in Argentina.

Einstein explained that in science there are two complementary goals: on the one hand is "the quest for an enlargement and enrichment of our understanding of some particular area of knowledge"; on the other hand, the one Einstein himself was interested in achieving, is "the endeavor to achieve a systematic unity of knowledge."

Although physics is regarded as an empirical science, Einstein remarked that there is no uniform method enabling a researcher to pass from experimental data to fundamental laws. The logical connection between principles and observed data was, for him, an "an act of intuition." He also pointed to the temporary character of fundamental laws of physics, as a yet unobserved fact can force a change in them; he also remarked that "experience is … the judge, but not the generator of fundamental laws."

Einstein observed that ideas or concepts, even if emanating from experience, have "a certain logical independence" and offered the notion of number in primitive societies as an example of concepts created by people without any scientific background. He insisted that there is no necessary road linking laws and facts from experience and, as an example, quoted Galileo's theorem on the proportionality between force and acceleration. Einstein remarked that it does not come "immediately from experience" it is, rather, a "free statement" coming from intuition. The history of mechanics before Galileo shows that it is not evident, even necessary. It is an arbitrary fact that, he stated, "the general theory of relativity has found it necessary to modify it."

For Einstein, not only fundamental laws proceed from "an act of imagination," the concept of acceleration is for him another example of a "free creation of the mind." He also remarked that laws can be falsified not only when a wrong consequence from them is found, or when they prove insufficiently general for the context in which they are defined, but also when some of its basic concepts cannot be expressed with the conceptual precision required to explain experimental facts. He indicated that this was the case regarding the concept of temperature in thermodynamics.

Nirenstein's 1925 paper essentially covers the text of *La Vida Literaria* we have summarized above. The final topic in Einstein *inédito* was presented as a question in Nirenstein's paper: "Is there an end to this development?" Einstein stated that physicists of his generation doubted it. The dialogue ended with the remark: "for us any theory contains as much truth as an equation can hold" attributed to Einstein; the latter actually expressed the view that "any theory is true only in the sense in which a parable can be true". He also omitted Einstein's notion that even if we cannot penetrate the ultimate truths, we nevertheless have the gratifying feeling that each generation of researchers will go deeper in their understanding than their predecessors.

**Final remarks**

In the two opening lectures delivered at the University of Buenos Aires, Einstein focused on the historical evolution of the concept of relativity and on the role played by delicate physical experiments in the emergence of his theory, leaving aside a brief epistemological analysis of the physical sciences that, we know now, he had intended to read in Buenos Aires. There were, as we have indicated above, several reasons for Einstein's decision to leave aside any topics that might have been regarded as controversial. One reason is that sections of the German-Argentine community objected to his visit on account of his pacifist attitude during the War. Such division was visible even in discussions at the academically minded German-Argentine cultural association before his visit and, during the visit, at the reception the German Embassy gave Einstein.

It is also true that the official atmosphere in Argentina at the time of Einstein's visit was also less receptive to his world views than it was at the time he was invited. By 1925 the university Reform movement was already in retreat and the progressive groups of intellectuals who supported his visit enthusiastically in 1922 (and continued to do so in 1925) were then beginning to lose their official influence. Nicolai, who criticized his university colleagues' fragility, paid dearly for the indiscretion. Traditional social groups, keen to entertain lesser visiting luminaries, were politely indifferent towards the eminent physicist. For some of the sponsors of Einstein's visit, his friendly, open, and wide contact with the Argentina press was, perhaps, not as much in tune with the new realities of official Argentina as Einstein may have thought; his openness clearly surprised some.

It is possible to argue that Nirenstein, a man of gentle disposition, with an aura of *bohemia*, may have made an effort to steer the visit in a direction that some groups of sponsors may have perceived as more in tune with the results they expected. When Nirenstein abstracted large sections of the *inédito* in his *Verbvm* paper he presented them as parts of a casual conversation sustained while he was driving his car. He did not disclose Einstein's original intention, which was nothing less than to read these sections at the formal grand opening of his series of lectures on the theory of relativity at the University of Buenos Aires, an occasion which was naturally to be attended by a wide and most influential audience. The audience, however, may have been more interested in a general philosophical panorama than in detailed physical experiments.

If the considerations sketched above played a role in Einstein's exchanges outside the field of theoretical physics while in Argentina, as we believe they did, they perhaps reflect excessive caution. There was little justification to keep his epistemological notes away from his academic audience and, after that, to keep them unpublished for six years after the visit had taken place. The views expressed in it did not fully contradict, but perhaps reinforced, those of Alberini and other leaders of the anti-positivist movement in Argentina. In fact, Einstein remained in some contact with Alberini after his 1925 trip. Close to 1930 he helped to get a short monograph of his, on the reception of contemporaneous German philosophy in Argentina, accepted for publication in Germany (Alberini 1930). However, after 1933, Einstein and Alberini's views on the political path Germany was beginning to follow were not necessarily convergent; there was no further contact from either side.

No doubt the existence of this manuscript of Einstein was known to a group of people before 1931. It is surprising that no full translation was considered by the editors of any of the philosophical or scientific periodicals published then in Argentina, or by official publications such as *Anales de la Universidad de Buenos Aires*, where Alberini and Nirenstein had considerable influence. It was only an *avant-gard* literary journal at the fringe of Argentine official intellectual life that had the sensitivity to translate the manuscript and make it public.

It appears that by 1931, at least to the eyes of academics in Argentina, the image of Einstein as a world-thinker was competing favorably with his image as a theoretical physicist. It has to be admitted that, by 1931, relativity theory was not as much at the top of the international scientific agenda as it had been before 1925. It has been remarked elsewhere (Eisenstaedt 1986) that the number of books on the theory of relativity published over the two decades immediately after the visit shrank to a few dozen.

Finally, the reader may have suspected, rightly, that the advocates of the *inédito*, Samuel Glusberg and Enrique Espinosa (Espinoza), who had it translated and published in 1931 and also reminded readers of Argentine publications in 1934 and 1955 of its existence, were one and the same person.[13] On occasion, each one of them lauded the achievements of the other.

**The Fate of Einstein's *inédito***

Einstein's *inédito* has been preserved, but is not in the public domain. It was sold through the offices of the well known German antiquarian J. A. Stargardt on June 6-7, 1978. In the auction's catalog[14] the manuscript is described as "Eigenh. [sc. eigenhändiges] Manuskript eines Vortrags über die Grundlagen der modernen Physik," presented in modified form by Einstein at the University of Buenos Aires in March 1925 which, as we know, was not the case. Einstein did not read this lecture in Argentina. In March 2006, the manuscript was sold again for 34,000 Euros to a private collector who wished to remain anonymous.[15] The 1978 auction catalog reproduces the first page of the manuscript and mentions that "the manuscript is in parts slightly discolored with brown stains" (see fig. 5 in the Appendix). Basically the same information is given in the catalog of 2006. Comparing the facsimile of the latter (see fig. 3 in the Appendix) to the photograph reproduced in 1931 (see fig. 2 in the Appendix), we also see that a vertical folding became visible over time. In the catalog of 2006, Albert Einstein's son, Hans Albert Einstein, certified the authenticity of the manuscript.

**Acknowledgments**

We thank the anonymous referees for their careful reading and useful suggestions; the editors of *Science in Context*, Leo Corry and Alexandre Métraux, for their interest in our work and also for their valuable advice; Barbara Wolf, at the Einstein Archives in Jerusalem, for informing us that the manuscript of the *inédito* had been sold in the antiquarian market; Wolfgang Mecklenburg, managing director of the "Autographenhandlung," J. A. Stargardt Ltd., Berlin, for kindly providing us with photographs of the pages of the original manuscript reproduced in their catalogs, and for details of successive auctions of the *inédito*; Christoph Hoffman, Berlin, for providing us with excellent scans of the Einstein documents from Stargardt's catalog 683. We also thank the linguist Jochen Dehmel, London, for his help with difficult passages in the German text. Finally, we thank Roberto Ferrari, Liliana Maghenzani, and Washington Pereyra for giving us access, at various times, to their important collections of documents. The Royal Society, London, contributed with a grant to support the research of one of us.

---

[13] The pseudonym is a happy fusion of the given name of poet Heinrich Heine, and the family name of philosopher Benedictus de Spinoza, the two men Glusberg said he admired most.
[14] *Katalog* 615, 1978, item 389, 107
[15] *Katalog* 683, 2006, item 425, 188.